\documentclass[preprint,aps]{revtex4}
\usepackage{graphicx}
\usepackage{dcolumn}
\usepackage{bm}

\begin{document}

\title{\bf{Self consistent theory of unipolar charge-carrier injection in metal/insulator/metal systems}}
\author{F. Neumann, Y. A. Genenko, C. Melzer and H. von Seggern} 
\affiliation{Institut f\"{u}r Materialwissenschaft, Technische Universit\"{a}t Darmstadt, 
Petersenstrasse 23, D-64287 Darmstadt, Germany}

\begin{abstract}
A consistent device model to describe current-voltage characteristics of metal/insulator/metal systems is developed. 
In this model the insulator and the metal electrodes are described within the same theoretical framework by using density 
of states  distributions. This approach leads to differential equations for the electric field which have to be 
solved in a self consistent manner by considering the continuity of the electric displacement and the electrochemical 
potential in the complete system. The model is capable of describing the current-voltage characteristics of
the metal/insulator/metal system in forward and reverse bias for arbitrary values of the metal/ insulator injection 
barriers. In the case of high injection barriers, approximations are provided offering a tool for comparison with 
experiments. Numerical calculations are performed exemplary using a simplified model of an organic semiconductor.
\end{abstract}

\maketitle

\section{Introduction}

Injection is an important factor in many electronic devices, especially in applications with insulators, where virtually all charge 
carriers have to be injected from the electrodes. Examples that have been the subject of interest are electronic devices built with 
organic semiconductors such as organic-light-emitting-diodes (OLEDs). Organic semiconductors show many properties of dielectric 
materials like relatively large band-gaps or the absence of intrinsic charge carriers. For the processes describing the conduction in 
organic semiconductors itself, sophisticated models were recently developed 
\cite{Scott1999,Malliaras1999,Arkhipov1998,Arkhipov1999,Arkhipov2003,Scott2003,Preezant2003,Shen2004,Woudenbergh2005,Novikov2006}. 
The proposed injection mechanisms are taking into account the stochastic, hopping character of the transport, the field-dependence of 
the mobility of charge carriers, the roughness-induced energetic disorder at the interface and the specific density of states (DOS) 
characteristic of these materials. However, the space charge effects were accounted so far only concerning the transport inside the 
organic materials while the description of injection itself was based on the classic works of Richardson and Fowler-Nordheim 
\cite{Lambert,Sze} considering injection as a single electron process. A comprehensive numerical model was developed by Tuti\v{s} et al. 
\cite{Tutis2001} comprising the hopping transport in single- and bilayer devices and tunnel injection from 
electrodes accounting for space-charge effect. However, besides the injection barriers given by the bare difference of the chemical 
potential in the metal and the lowest occupied orbital (LUMO) in organic material this model involves a range of artifacts like 
tunneling factors, effective attempt frequencies etc. Being a very sofisticated numerical tool and presenting good agreement with 
experiments this approach does not allow analytical fitting of current-voltage characteristics which gives insight in major mechanisms
controlling injection.

In this work we develop a device model for a metal/insulator/metal system including the description of charge carrier injection at the 
metal/insulator interfaces. This injection model takes into account the electrostatic potential generated by the charges injected. 
The treatment of this problem faces the problem of self consistency, since the amount of injected charges depends on the height of the 
injection barrier, while the electrostatic potential generated by this charge modifies the height of the injection barrier itself. 
This problem is solved by defining the boundary conditions far away from the interface, where the influence of the interface can be 
ignored. As a result, charge carrier densities and electric field distributions in the respective medias are coupled and depend on 
the conditions of the system. By describing the electrochemical potential as well as the dielectric displacement continuously all 
over the entire system the electric field and charge-carrier distributions can be calculated. One can model current-voltage-(IV) 
characteristics in which either charge injection or the transport through the insulating layer determines the current. The model 
includes barrier lowering, but this lowering differs significantly from the lowering due to the image charge potential which is a 
direct consequence of the one electron picture. 

The problem considered in this paper includes the presence of both, injecting and ejecting electrode with an organic semiconductor in 
between and, hence, a built-in potential in the system. Both interfaces are considered self-consistently involving no additional 
characteristics but the bare injection barriers given by the difference of the chemical potentials in the metals and the LUMO. 
Electric field distributions in the device are calculated analytically for different barriers heights in equilibrium. Then the field 
distributions are calculated numerically in the presence of a steady-state current. It is shown that these field distributions are 
qualitatively different for injection barriers smaller and larger than some characteristic value. Using the known solution for the 
electric field at a fixed current the current-voltage characteristics are computed. They exhibit, depending on the barrier heights and bias, 
distinct areas of linear, exponential and quadratic dependence. In the case of devices with large barriers an analytic approximation 
for the field distribution is derived which describes the numerical results with high accuracy. The obtained approximation is similar 
to the well-known Mott-barrier formula but includes an effective barrier height which results from the consistent treatment of the 
metal/dielectric interface.

\section{The model}

Let us consider an insulator of thickness $L$ sandwiched in between two metal electrodes. The insulator is supposed to be extended 
over
the space with $-L/2<x<L/2$, whereas the metal electrodes are extended over the semi spaces with $x<-L/2$ and $x>L/2$, respectively. 
In the following sections the theoretical model describing the insulator and the metal electrodes is introduced. The description is 
based on the specific density of states (DOS) distributions of the employed media. Differential equations will be derived, describing 
the electric field distribution by the requirement of local chemical equilibrium.

\subsection{The Electrodes}

Assuming the free electron model and the Thomas-Fermi approximation \cite{Ashcroft}, the electrochemical potentials $\kappa^\pm_m$ of 
the metals as a function of the spatial coordinate $x$ read,
\begin{equation}
\kappa^\pm_m(x)=\frac{\hbar^2}{2m^\pm}(3\pi^2n^\pm_m(x))^{2/3}-e\phi(x) + E_b^\pm,
\label{kappametal}
\end{equation}
Here $n^\pm_m(x)$ are the electron densities, $e$ is the elementary charge and $m^\pm$ the effective electron masses in the metals, 
$\phi(x)$ the electrostatic potential and $E_b^\pm$ the bottom of the conduction bands. All quantities with indices $\pm$ are assigned 
to the electrode on the right ($x>L/2$) and the left hand side ($x<-L/2$), respectively.
The steady state current density $j$ is given by the conductivity $\sigma$ and the derivative of $\kappa(x)$ \cite{Landau}. Since we 
consider a one-dimensional, monopolar case, the current remains constant across the whole space,
\begin{equation}
j=\frac{\sigma}{e}\frac{d\kappa(x)}{dx}=\mbox{const}.
\label{currentgeneral}
\end{equation}
The conductivities of the metals are proportional to the electron mobilities $\mu^\pm_m$ and the charge-carrier densities far away from 
the contacts, $n^\pm_\infty$: $\sigma^\pm_m=e \mu^\pm_m n^\pm_\infty$.
Due to the charge transfer between the metals and the insulator, space charge regions emerge, which modify the electric fields 
$F^\pm_m(x)$ also in the metal. According to Gauss's law, the derivatives of the electric field are proportional to the excess electron 
densities $\delta n^\pm(x)=n^\pm_m(x)-n^\pm_\infty $:
\begin{equation}
F^{\pm'}_m(x)=-\frac{e}{\epsilon_0}\delta n^\pm (x).
\label{gaussmetal}
\end{equation}
where $\epsilon_0$ is the permittivity of free space. 

Since electron densities in metals are rather high, the value for the excess 
electron densities is small in comparison with the background electron densities, $|\delta n^\pm(x)|\ll n^\pm_\infty$. Assuming the 
linear Thomas-Fermi approximation a straightforward calculation leads to differential equations for $F^\pm_m$, which read:
\begin{equation}
\frac{j}{e\mu^\pm_mn^\pm_\infty}=-\frac{2\epsilon_0\kappa^\pm_\infty}{3e^2n_\infty}F^{\pm''}_m(x) + F^\pm_m(x).
\label{dglmetal}
\end{equation}
Here $\kappa^\pm_\infty$ are the chemical potentials in the metals at an infinite distance from the metal/insulator interfaces.
There space charge effects vanish and hence gradients of $F^\pm_m(x)$. With this boundary condition the solutions for $F^\pm_m(x)$ 
read,
\begin{equation}
F^\pm_m(x)=\left [ F^\pm_m(\pm L/2)-\frac{j}{\sigma^\pm_{m}} \right ] \exp \left [ \mp \frac{x\mp L/2}{l^\pm_{TF}} \right ] + 
\frac{j}{\sigma^\pm_{m}},
\label{solutionmetal}
\end{equation}
with
\begin{equation}
l^\pm_{TF}=\sqrt{\frac{2}{3}\frac{\epsilon_0\kappa^\pm_\infty}{e^2n^\pm_\infty}},
\label{ld}
\end{equation}
being the Thomas-Fermi lengths, defining the typical length scales of the metals ($l_{TF}\simeq 10^{-10}m$). It is clear from 
expressions (\ref{solutionmetal}) that our assumption of the metal semi-spaces is, in fact, not necessary. With the very short 
length $l_{TF}$ the results are valid for any experimental metal thickness.
In Eq. (\ref{solutionmetal}) the electric fields in the metals at the metal/insulator interfaces $F^\pm_m(\pm L/2)$ remain as the 
only unknown quantities.

\subsection{The Insulator}

The energetic difference between the electrochemical potentials in the metal electrodes, at an infinite distance from the contact, 
and the bottom of the conduction band in the insulator is defined as the injection barrier $\Delta^\pm$. Accordingly the energetic 
difference of the $\Delta^\pm$s is related to the difference between the electrode work functions $E^\pm_A$,
\begin{equation}
E_A^- - \Delta^-  = E_A^+ - \Delta^+.
\label{deltadelta}
\end{equation}
The introduction of a bottom of the conduction band in the insulator means that one can calculate charge carrier densities by using 
Boltzmann statistics.
\begin{equation}
n_s(x)=\int\limits_{-\infty}^\infty g_s(E-\Delta^- - \kappa_\infty^-)\exp \left ( \frac{\kappa_s(x)+e\phi(x)-E}{kT} \right ) dE,
\label{nsboltzmann}
\end{equation}
Here, $g_s(E)$ is the density of states of the insulator with $g_s(E<0)=0$, $T$ is the absolute temperature and $k$ is the Boltzmann 
constant. The energy scale, $E$, has been adjusted to the bottom of the conduction band in the left metal electrode. Thus, the 
electrochemical potential $\kappa_{s}$ can be expressed in terms of the charge-carrier density $n_s$,
\begin{equation}
\kappa_{s}(x) = kT\ln \left ( \frac{n_s(x)}{{\cal N}} \right ) +\Delta^-+\kappa^-_\infty-e\phi(x),
\label{kappaorganic}
\end{equation}
where ${\cal N}$ is defined by
\begin{equation}
{\cal N}=\int\limits_{-\infty}^\infty g_s(E)\exp(-E/kT) dE
\label{neff}
\end{equation}
and can be understood as the effective total density of states available in the insulator at a given temperature $T$. 
One should realize that the $T$ dependence of ${\cal N}$ becomes weak in the case of a narrow-band material.

For wide band gap insulators all charge-carriers in the extended states, $n_s$, are excess charge-carriers appearing 
in Eq.(\ref{kappaorganic}). They have to be considered in Gauss's law,
\begin{equation}
F_s'(x)=-\frac{e}{\epsilon\epsilon_0}n_s(x),
\label{gaussorganic}
\end{equation}
where $\epsilon$ is the relative permittivity of the insulator. With Eqs.(\ref{currentgeneral}), (\ref{kappaorganic}), 
(\ref{gaussorganic}) and $\sigma_s(x)=e\mu_s n_s(x)$ a differential equation for the electric field in the insulator, $F_s$, 
is obtained,
\begin{equation}
\frac{j}{\mu_s\epsilon\epsilon_0}=-\frac{kT}{e}F_s''(x)-F_s(x)F_s'(x),
\label{dglorganic}
\end{equation}
where $\mu_s$ and $\sigma_s$ are the electron mobility and conductivity in the insulator, respectively. Alternatively, 
this differential equation could  be derived applying the drift-diffusion model and the Einstein relation, which relates 
the mobility and the diffusivity in non-degenerate systems \cite{Ashcroft,Neumann06Ein}.

\subsection{Boundary conditions and self-consistency}

The self consistent treatment of charge transport through an insulator sandwiched between two metals requires continuity 
of the electrical displacement and of the electrochemical potential across the whole system. In particular one may write:
\begin{eqnarray}
\kappa(x)& = & continuous,
\label{bcgeneralkappa} \\
\epsilon F(x)& = & continuous.
\label{bcgeneralfield}
\end{eqnarray}
These conditions have to be fulfilled particularly at the metal/insulator interfaces, eliminating the unknown integration constants. 
It follows from Eq.(\ref{bcgeneralfield}) that the electric fields in the metals and the insulator at the interface are related 
by: $F_s(\pm L/2)=F^\pm_m(\pm L/2)/\epsilon$. From Eq.(\ref{bcgeneralkappa}) nontrivial boundary conditions for Eq.(\ref{dglorganic}) 
are obtained:
\begin{equation}
\ln(-\frac{\epsilon\epsilon_0}{e{\cal N}}F'_s(\pm L/2)) + 
\frac{\Delta^\pm}{kT}\mp \epsilon \frac{el^\pm_{TF}}{kT}F_s(\pm L/2) = 
\mp \frac{l^\pm_{TF}}{\mu^\pm_m n^\pm_\infty kT}j,
\label{bcss}
\end{equation}
\noindent which depend on parameters of both the insulator and the metal electrodes.

Indeed, these boundary conditions are virtually independent of the net current density. In Eqs.(\ref{bcss}) the current density 
$j$ is multiplied by small factors $l^\pm_{TF}/\mu_m^\pm n_\infty^\pm kT$ and hence, the boundary conditions are independent of 
$j$ in most practical cases. Under this approximation the density of injected electrons at the interfaces reads,
\begin{equation}
n_s(\pm L/2)={\cal N} \exp \left [ - \Delta^\pm/kT\pm \epsilon \frac{el_{TF}^\pm}{kT}F_s(\pm L/2) \right ]. 
\label{nnull}
\end{equation}
Hence, the dependence of $F_s(\pm L/2)$ or $n_s(\pm L/2)$ on the current is mainly due to Eq.(\ref{dglorganic}). From Eq.(\ref{nnull}) 
it is evident that the height of the injection barrier is modified by the electric field at the interface. This leads to the definition
 of an effective injection barrier,
\begin{equation}
\Delta_{eff}^\pm=\Delta^\pm  \mp \epsilon e l_{TF}^\pm F_s(\pm L/2).
\label{deltaeff}
\end{equation}
The barrier modification is a direct consequence of the charge transfer from the metal to the insulator and corresponds to the 
potential energy, electrons gain or lose in the electric field of the metal.

\section{Physical and numerical analysis}

We now show the results of analytical and numerical calculations. The section is organized as follows: In the first part we show 
results for calculations in equilibrium where charge carriers diffuse from the metal electrodes into the insulator. In the second 
part we calculate electric field distributions in steady state and finally we analyze the resulting current-voltage (IV) 
characteristics.

First, the material parameter of the metal electrodes and the insulator will be specified.
As mentioned in the introduction we chose a simplified model of an organic semiconductor as example for an insulator. These 
materials are characterized by band-gap energies ranging form $1$ to $3$ eV making thermal excitation to conduction states virtually 
impossible. As in insulators, charge transport is therefore dominated by excess charge carriers. The DOS of organic semiconductors 
is believed to have a gaussian shape \cite{Baessler1993}, impeding in general the applicability of Eq.(\ref{nsboltzmann}) and 
requiring the use of Fermi-statistics. However, in the case of weak disorder charge carrier trapping in tail states of the 
DOS is negligible and Boltzmann-statistics can be applied. We assume that the material specific parameters for the organic 
semiconductor adopt the typical values of:
\begin{equation}
{\cal N}=10^{21}\mbox{cm}^{-3} \quad \mbox{at} \quad T=300\mbox{ K}, \quad \epsilon=3 \quad \mbox{and} \quad \mu_s=10^{-4}\mbox{cm}^{2}/Vs.
\label{parorg}
\end{equation}
An organic semiconductor with a thickness of $L=100 \mbox{ nm}$ will be considered, typical for applications in OLEDs. As examples for 
the metal electrodes, calcium, barium and magnesium are chosen. These metals are characterized by their electron densities in the 
conduction band $n_\infty$, their chemical potentials at infinite distance from the contact $\kappa_\infty$, which at $T=0$ equal 
their Fermi-energies, their electron mobilities $\mu_m$ and their electron work functions $E_A$. The values for the material 
parameters are summarized in Tab.\ref{metalpar}.
Here, the barriers $\Delta^\pm$ are given by the energetic difference between the transport level in the organic semiconductor and 
$\kappa^\pm_\infty$ of the respective metal electrode.

\begin{table}[h]
    \begin{center}
      \caption{{\itshape Parameter values \cite{Ashcroft} for the different metal electrodes considered in this work.}}
        \begin{tabular}[t]{|c||c|c|c|c|} \hline
 & $n_\infty $ [$10^{22}cm^{-3}$] & $\kappa_\infty$ [\mbox{ eV}] &$\mu_m$ [$\mbox{cm}^{2}/Vs$] &$E_A [\mbox{ eV}]$    \\ \hline
 Ca& 2.6 & 4.68 & 66.7 & 2.87  \\ \hline
  Ba& 3.2 & 3.65 & 5.07 & 2.7  \\ \hline
 Mg & 8.6 & 7.13 & 16.9 & 3.68  \\ \hline
        \end{tabular}
\label{metalpar}
    \end{center}
\end{table}

In the following the solutions for the equilibrium situation will be discussed, where space charge zones are formed due to diffusive 
charge-carrier transport from the electrodes into the insulator. Then a constant external voltage will be considered, driving a 
constant current through the insulator, and the calculated IV-characteristics which are crucial for comparison with experiment will 
be presented.

\subsection{Equilibrium} 

In equilibrium, the current density $j$ vanishes and Eq.(\ref{dglorganic}) can be integrated:
\begin{equation}
\frac{kT}{e}F_s'(x)+{1\over 2} F_s^2(x)=-2 \left ( \frac{kT}{e}\right )^2 \Lambda^2,
\label{dglorglambda}
\end{equation}
where $\Lambda$ is a constant. The derivative of the electric field has to be negative due to Eq.(\ref{gaussorganic}). At some 
position between the contacts the electric field vanishes, thus, $\Lambda^2>0$ and an analytic solution for the electric field 
can be found,
\begin{equation}
F_s(x)=-\frac{2kT}{e}\Lambda \tan \left( \Lambda x + \lambda \right ),
\label{orgequifin}
\end{equation}
with $\lambda$ being the integration constant. This solution describes the following processes: charge carriers are injected from 
both metal electrodes and diffuse far into the bulk of the organic semiconductor. The resulting space charge generates an electric 
field compensating - in equilibrium - the diffusion current. While the widths of the space charge zones in the metals are of the 
order of $l_{TF}$, they can be extended over the whole insulating layer. This means, that at some point the diffusion currents 
caused by electrons injected from both electrodes compensate each other: here the electric field is zero. 

\begin{figure}[h]
\begin{center}
\includegraphics[height=6.4cm,width=7.5cm]{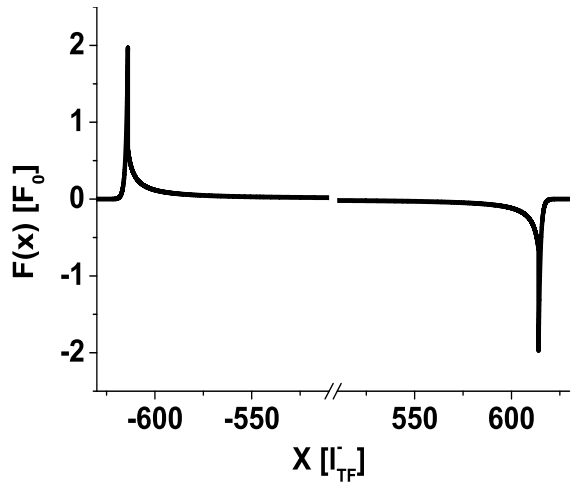}
\caption{{\itshape \small Distribution of the electric field $F$ in units of $F_0=kT/el_{TF}^-$ as a function of the coordinate $x$ 
in units of $l_{TF}^-$ in case of equilibrium ($j=0$) and barrier free injection ($\Delta^-=\Delta^+=0$) from two Ca electrodes. 
Notice that the $x$-axis is extended between $-L/2l_{TF}^-$ and $+L/2l_{TF}^- \simeq 610$ and interupted between $x=-490$ and 
$x=+490$. }}
\label{finequid0caca}
\end{center}
\end{figure}

From the analytical solution of Eq.(\ref{dglorglambda}), the boundary conditions (\ref{bcss}) can be simplified to a nonlinear system 
of equations determining the integration constants $\Lambda$ and $\lambda$: 
\begin{equation}
\ln \left ( \frac{\epsilon\epsilon_0}{{\cal N}}\frac{2kT\Lambda^2}{e^2}\frac{1}{\cos^2(\pm\Lambda L/2 + \lambda)} \right ) + 
\frac{\Delta^\pm}{kT}\pm 2\epsilon\Lambda l^\pm_{TF} \tan(\pm\Lambda L/2 +\lambda)  = 0
\label{bcequifin}
\end{equation}
In general, Eqs.(\ref{bcequifin}) have to be solved 
numerically. However, for the situation of symmetric electrodes the solution 
for the electric field is antisymmetric with respect to $x=0$, which means $\lambda=0$. In Fig.\ref{finequid0caca}
we show the solution for the electric field of an organic semiconductor sandwiched between two calcium electrodes. The injection is 
supposed to be barrier free ($\Delta^-=\Delta^+=0$). The introduction of nonvanishing injection barriers leads to a reduction of the electric field at the interface, although the 
dependence of the electric field on the coordinate is still given by Eq.(\ref{orgequifin}).

\begin{figure}[h]
\begin{center}
\includegraphics[height=6.4cm,width=7.5cm]{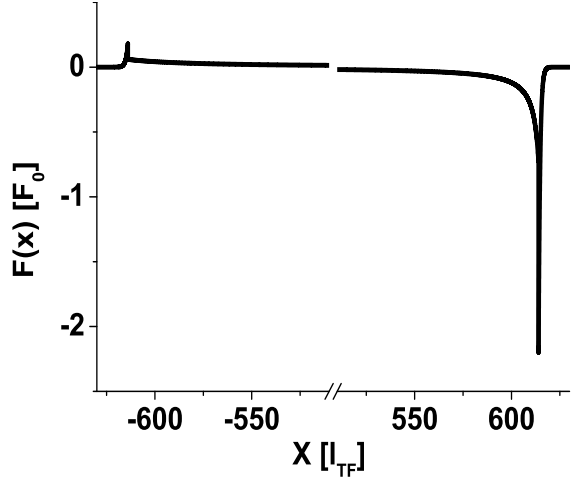}
\caption{{\itshape \small Distribution of the electric field $F$ in units of $F_0=kT/el_{TF}^-$ as a function of the coordinate 
$x$ in units of $l_{TF}^-$ in equilibrium ($j=0$) with an injection barrier of $\Delta^-=0.17\mbox{ eV}$ for a Ca electrode at 
$x=-L/2$ 
and a vanishing injection barrier for the barium electrode at $x=+L/2$. Notice that the $x$-axis is  interupted between $x=-490$ 
and $x=+490$. }}
\label{finequicaba}
\end{center}
\end{figure}

Now the electrode at $x=+L/2$ will be changed from calcium to barium, to demonstrate the consequences of an unbalanced 
charge-carrier injection.
The injection process at the barium side of the device at $x=+L/2$ is supposed to be barrier free ($\Delta^+=0$), the injection 
barrier for the calcium/organic interface at $x=-L/2$ is then calculated with the help of Eq.(\ref{deltadelta}) and reaches the 
value of $\Delta^-=0.17 \mbox{ eV}$. The effect of an unbalanced charge-carrier injection can be seen in Fig.\ref{finequicaba}. 
A significant asymmetry in the computed distribution of the electric field can be observed. 

Integration of the electric field over the whole system results in the built-in voltage $V_{BI}$,
\begin{equation}
-V_{BI}=\int\limits_{-\infty}^\infty F(x) dx = F_s(-L/2)\,\epsilon\, l_{TF}^- + \int\limits_{-L/2}^{L/2} F_s(x) dx + 
F_s(L/2)\,\epsilon\, l_{TF}^+ 
\label{VBIintegral}
\end{equation}
\noindent using Eqs.(\ref{solutionmetal}) and (\ref{bcgeneralfield}). It looks as if the integration runs virtually over the effective
length of the device 
\begin{equation}
L_{eff}=L+\epsilon (l_{TF}^- +   l_{TF}^+).
\label{Leff} 
\end{equation}
\noindent This length will be used later by the evaluation of the voltage on the device in a steady-state. The integral in 
Eq.(\ref{VBIintegral}) can be calculated analytically using the solution (\ref{orgequifin}) and boundary conditions (\ref{bcequifin}):
\begin{equation}
eV_{BI}=E_A^- - E_A^+
\label{VBI}
\end{equation}
and is independent of the values for the integration constants $\Lambda$ and $\lambda$. One should be aware of that the built-in 
potential drops over the insulator and the metal electrodes.

\subsection{Steady state}

Now, the steady state situation is considered, where a constant current $j$ flows  through the system. 
In this case, the differential equation (\ref{dglorganic}), describing the 
electric field distribution in the organic media has to be solved numerically with respect to the boundary conditions defined by 
Eqs.(\ref{bcss}).

\begin{figure}[h]
\begin{center}
\includegraphics[height=6.4cm,width=7.5cm]{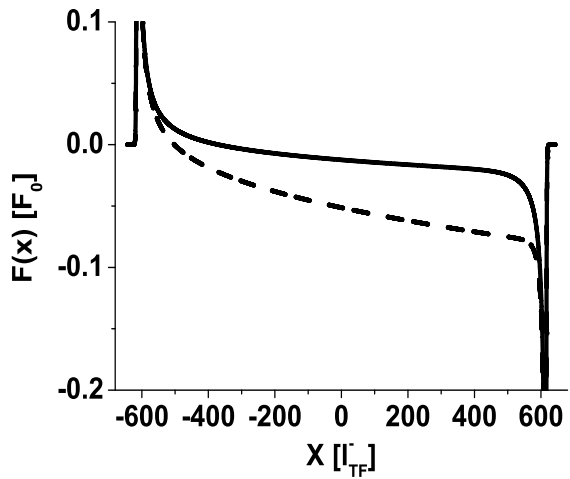}
\caption{{\itshape \small Distribution of the electric field $F$ in units of $F_0=kT/el_{TF}^-$ in steady state for current 
densities of $|j|=10\mbox{ mA}/\mbox{cm}^{2}$(solid line) and $100\mbox{ mA}/\mbox{cm}^{2}$(dashed line) as a function of the coordinate $x$ in units of $l_{TF}^-$. 
The charge injection is supposed to be barrier free with calcium electrodes on both interfaces. We comment that the maxima of the 
electric field near the interfaces are cut off in this plot and do not change in value with respect to the equilibrium.}}
\label{finssd0caca}
\end{center}
\end{figure}

The solution for the electric field of an organic semiconductor sandwiched between two calcium electrodes is shown in 
Fig.\ref{finssd0caca}. 
The calculation was carried out for barrier free injection at both metal/organic interfaces 
($\Delta^-=\Delta^+=0$) and steady state current densities of $|j|=10\frac{\mbox{ mA}}{\mbox{cm}^{2}}$ and $100\frac{\mbox{ mA}}{\mbox{cm}^{2}}$.
In case of vanishing injection barriers many charge-carriers are injected into the organic semiconductor. This requires strong 
electric fields to compensate the diffusion currents near the interfaces to such an extend that the net current is $j$. This 
field is positive on the injecting electrode and negative on the ejecting electrode. Thus, there has to be a position $x_0$, 
often referred to as the virtual electrode \cite{Rose1955}, where the electric field changes sign. When the current increases, 
this position shifts towards the injecting electrode. It can be seen from Fig.\ref{finssd0caca} that $x_0$  shifts from 
approximately $25 \mbox{ nm}$ to $10 \mbox{ nm}$ once $|j|$ is increased from $10$ to $100\frac{\mbox{ mA}}{\mbox{cm}^{2}}$.
In between the two electrodes, the electric field of the barrier free case follows approximately the electric field of a space 
charge limited system assuming ohmic boundary conditions \cite{Lambert}. In this situation the dipole layer at the left electrode 
acts as a source of charge-carriers, the right electrode can be understood as a sink.

\begin{figure}[h]
\begin{center}
\includegraphics[height=6.4cm,width=7.5cm]{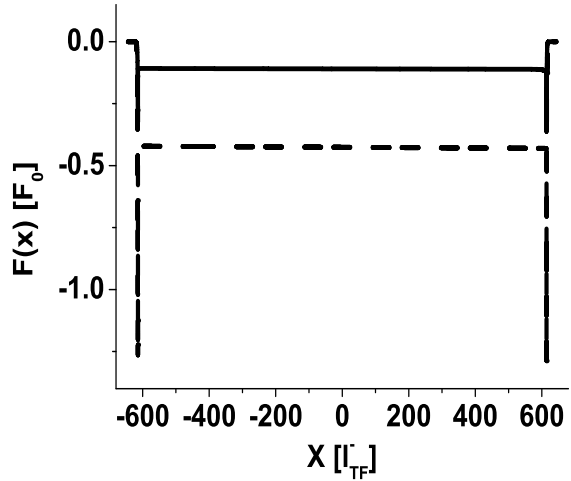}
\caption{{\itshape \small Distribution of the electric field $F$ in units of $F_0=kT/el_{TF}^-$ as a function of the coordinate 
$x$ in units of $l_{TF}^-$ in steady state for current densities of $|j|=10\mbox{ mA}/\mbox{cm}^{2}$ (solid line) and $100 \mbox{ mA}/\mbox{cm}^{2}$(dashed line). 
The organic layer of thickness $L=100\mbox{ nm}$ is contacted with calcium and barium with injection barriers of $\Delta^-=0.35\mbox{ eV}$ and 
$\Delta^+=0.18\mbox{ eV}$ respectively. }}
\label{finssd035caba}
\end{center}
\end{figure}

The introduction of non vanishing injection barriers reduces the amount of charge-carriers present in the organic semiconductor 
and therefore the importance of diffusion. 
When the injection barrier at $-L/2$ exceeds some critical value no compensation of 
diffusion is necessary and the electric field near the interface becomes negative. The charge-carrier reservoir is exhausted and 
no virtual electrode is formed. In Fig.\ref{finssd035caba}
the field distribution of a system is shown, where an organic layer 
with larger band gap forms a contact with calcium and barium electrodes. The injection barrier heights assigned to the contacts 
are $\Delta^-=0.35\mbox{ eV}$ and $\Delta^+=0.18\mbox{ eV}$, respectively. The electric field strength is virtually independent 
of $x$ while its value depends on the current density. The absence of extended space charge zones in the insulator indicates that 
the system is in an injection limited mode.

\subsection{Current-voltage-characteristics}

Knowledge about the distribution of the electric field allows one to determine the voltage drop $V$ for a given current density 
$j$ and hence to calculate the IV-characteristics. After solving the differential equations for the electric field for a given 
current density, the voltage applied to the system can easily be calculated by integration. For the steady-state case, however, 
one 
cannot define the voltage drop over the infinite  range  of integration as in Eq.(\ref{VBIintegral}) since the electric fields
in the metals, Eq.(\ref{solutionmetal}), are asymptotically constant and do not vanish. To compare with experimentally observed 
current-voltage characteristics, one has to account for the built-in voltage, Eq.(\ref{VBIintegral}), which drops virtually over 
the effective device length, Eq.(\ref{Leff}), and can define the voltage drop over the same length:     
\begin{equation}
V=-\int\limits_{-L/2-\epsilon l_{TF}^- }^{L/2+\epsilon l_{TF}^+} F(x) dx - V_{BI}
\label{voltagefin}
\end{equation}

\begin{figure}[h]
\begin{center}
\includegraphics[height=6.4cm,width=7.5cm]{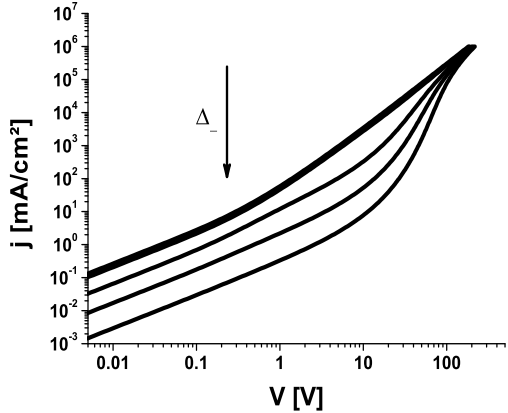}
\caption{{\itshape \small IV-characteristics for an organic layer with thickness $L=100\mbox{ nm}$ contacted with calcium electrodes on 
both interfaces and different injection barriers of $\Delta^-=\Delta^+=0;0.2;0.3;0.35;0.4\mbox{ eV}$.}}
\label{finivcacadmult}
\end{center}
\end{figure}

In Fig.\ref{finivcacadmult} we present IV-characteristics of a Ca/organic($100\mbox{ nm}$)/Ca system. The injection barriers are varied 
from $0\mbox{ eV}$ to $0.4\mbox{ eV}$.
The IV-curve calculated for vanishing injection barriers represents a limitation of the current flow through the system. At 
voltages up to $\sim 0.3 V$ the IV-characteristic shows an ohmic-like $j\sim V$ dependence. We emphasize that this ohmic behavior 
is not due to intrinsic charge-carriers but a result of charge carriers injected from the electrodes in the organic semiconductor 
even in equilibrium (Fig.\ref{finequid0caca}). Above $\sim 0.3 V$ the system shows the Mott-Gurney $j \sim V^2$ dependence of the 
current on the voltage. Here the system is obviously in a space charge limited mode. The IV-characteristic does not change 
considerably up to a barrier height of $\Delta_{\mbox{crit}}\simeq 0.27\mbox{ eV}$. The current flow is for all voltages limited by the 
transport through the bulk of the organic semiconductor and the injection process has only a secondary effect.
Introducing injection barriers $>0.27\mbox{ eV}$, the current decreases by orders of magnitude. Nevertheless, all IV-characteristics 
show the same $j\sim V$ ohmic-like dependence in the low voltage region like the curve calculated for barrier free injection. 
For high $\Delta$s this ohmic regime is followed by an exponential increase of the current on the voltage. For even higher 
voltages all curves approach the space charge limit and coincide with the barrier free IV-curve.

\begin{figure}[h]
\begin{center}
\includegraphics[height=6.4cm,width=7.5cm]{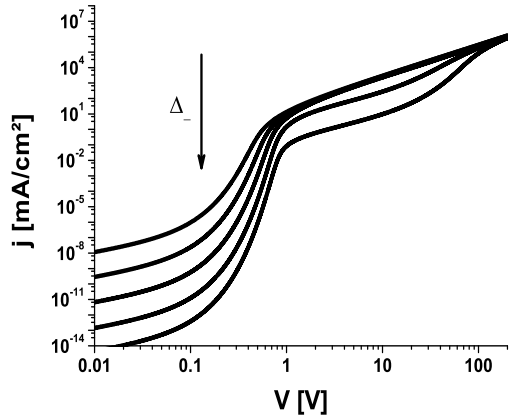}
\caption{{\itshape \small IV-characteristics for an organic layer with thickness $L=100\mbox{ nm}$ contacted with a calcium electrode 
at $x=-L/2$ and a magnesium electrode at $x=+L/2$. The injection barrier of the calcium electrode is varied from $0 \mbox{ eV}$ to 
$0.4 \mbox{ eV}$ with an increment of $0.1\mbox{ eV}$. The work function of the metals differs by an amount of $0.81\mbox{ eV}$, accordingly the 
injection barrier of the magnesium electrode is higher by that value.}}
\label{finivcamgdmult}
\end{center}
\end{figure}

In Fig.\ref{finivcamgdmult} we introduce a built-in voltage by changing the metal at $x=+L/2$ from calcium to magnesium. 
The difference in the work functions corresponds to a built-in voltage of $V_{BI}=-0.81\mbox{ eV}$. The injection barrier on the 
calcium electrode at $x=-L/2$ is varied from $0$ to $0.4 \mbox{ eV}$. Correspondingly, injection barriers on the magnesium 
electrode at $x=+L/2$ range from $0.81$ to $1.21\mbox{ eV}$. 

The computed IV-characteristics for high injection barriers can be understood by considering some simple approximations. In 
the case of high injection barriers the amount of injected charges is small and the electric field in the insulator can be 
conveniently decomposed into a constant mean value and a space charge part which is assumed to be comparably small,
\begin{equation}
F_s(x)=F_{s0} + f(x) \quad \quad \mbox{where} \quad \quad |f(x)| \ll |F_{s0}|.
\label{linfield}
\end{equation}
The mean value of the electric field is determined by the voltage drop on the effective length of the metal/insulator/metal system, 
(\ref{voltagefin}):
\begin{equation}
F_{s0}=-\frac{V + V_{BI}}{L_{eff}},
\label{linfs0}
\end{equation}
which implies that the integral of $f(x)$ over the system length equals zero. 

The decomposition of the electric field represented by Eq.(\ref{linfield}) leads to a linearized form of the differential equation 
for the electric field in the insulator:
\begin{equation}
  f''(x) + \frac{eF_{s0}}{kT}f'(x) + \frac{e j}{\mu_s \epsilon\epsilon_0 kT} = 0.
\label{lindgl}
\end{equation}
which should be supplemented by the linearized boundary conditions, Eqs.(\ref{nnull}). Taking into account that in the considered 
systems $l_{TF}^\pm/L $ are as small as $10^{-3}$, the solution of Eq. (\ref{lindgl}) reads
\begin{equation}
f(x)=B\left[\frac{2kT}{e F_{s0} L} \sinh\left( \frac{eF_{s0}L}{2kT}\right ) -\exp\left ( -\frac{eF_{s0}}{kT}x\right ) \right] - 
\frac{j}{ \mu_s \epsilon\epsilon_0 F_{s0}}x,
\label{linfx}
\end{equation}
here the quantity $B$ is an abbreviation of
\begin{equation}
B=\frac{{\cal N} kT}{2\epsilon\epsilon_0 F_{s0}\sinh(eF_{s0}L/2kT)}\left[\exp\left ( -\frac{\Delta_{eff}^+}{kT} \right ) 
-\exp \left ( -\frac{\Delta_{eff}^-}{kT} \right )\right].
\label{deltaexp}
\end{equation}
where we assume, approximately,
\begin{equation}
\Delta_{eff}^\pm \simeq \Delta^\pm  \mp \epsilon e l_{TF}^\pm F_{s0}.
\label{deltaeffapprox}
\end{equation}

\begin{figure}[h]
\begin{center}
\includegraphics[height=6.4cm,width=7.5cm]{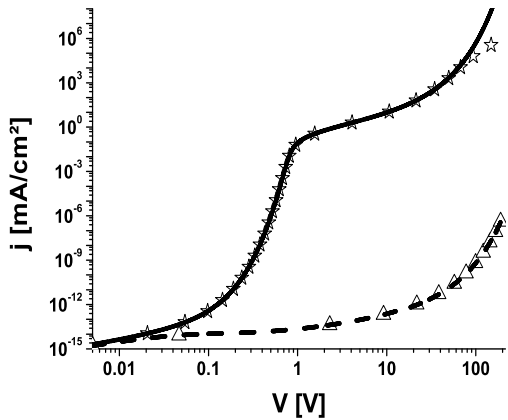}
\caption{{\itshape \small IV-characteristics for an organic layer with thickness $L=100\mbox{ nm}$ contacted with a calcium electrode 
at $x=-L/2$ and a magnesium electrode at $x=+L/2$. The stars and the triangles correspond to the the IV-characteristic calculated 
for forward and reverse bias, while the solid and dashed lines are the corresponding approximate solutions given by 
Eq.(\ref{ivapprox}). }}
\label{finivcamgfr}
\end{center}
\end{figure}

The analytic solution results directly in the IV-characteristic of a system with high injection barriers:
\begin{equation}
j=\frac{e{\cal N} \mu_s F_{s0}}{2\sinh(eF_{s0}L/2kT)} \left[ \exp \left ( \frac{eF_{s0}L}{2kT} - 
\frac{\Delta^+_{eff}}{kT}\right ) - \exp\left ( -\frac{eF_{s0}L}{2kT} - \frac{\Delta^-_{eff}}{kT}\right ) \right ].
\label{ivapproxsym}
\end{equation}
This relation is symmetric in the injection barriers and is capable of describing the current flow in both possible directions. 
In the following we focus on electron transport from the left to the right electrode. For such situations it is convenient to 
examine a nonsymmetric form of Eq.(\ref{ivapproxsym}) which reads:
\begin{equation}
j=-e\mu_s \frac{V+V_{BI}}{L_{eff}} {\cal N} \exp \left ( -\frac{\Delta_{eff}^-}{kT} \right ) 
\frac{\exp\left( -\frac{eV}{kT}\right ) -1}{\exp \left ( -\frac{e(V+V_{BI})}{kT}\frac{L}{L_{eff}} \right) -1}.
\label{ivapprox}
\end{equation}
Equation (\ref{ivapprox}) resembles the current-voltage characteristic for Mott-barriers \cite{Sze}. An essential difference 
involves the earlier introduced barrier lowering in the electrodes. In Fig.\ref{finivcamgfr} 
we compare the approximate solution 
of the injection problem, Eq.(\ref{ivapprox}), with the numerical solution. The agreement is perfect for reverse and forward bias 
until the space charge effects become dominant and Eq.(\ref{lindgl}) loses its validity.

\begin{figure}[h]
\begin{center}
\includegraphics[height=6.4cm,width=7.5cm]{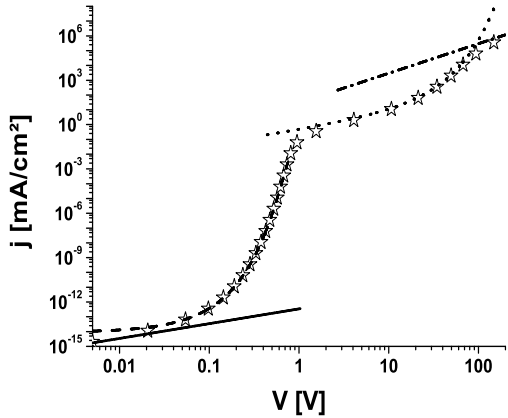}
\caption{{\itshape \small IV-characteristics for an organic layer with thickness $L=100\mbox{ nm}$ contacted with a calcium electrode at 
$x=-L/2$ and a magnesium electrode at $x=+L/2$. The stars represent the exact numerically calculated IV-characteristic for forward 
bias. The solid line corresponds to the linear part approximated by Eq.(\ref{jappr1}), the dashed line to the exponential current 
increase given by Eq.(\ref{jappr2}), the dotted line to the injection current described by Eq.(\ref{jappr3}) and the dashed dotted 
line to a SCLC given by $|j|=(9/8)\epsilon\epsilon_0 \mu V^2/L^3$. }}
\label{finivcamgapprox}
\end{center}
\end{figure}

In Fig.\ref{finivcamgapprox} we compare different regimes of Eq.(\ref{ivapprox}) with the exact numerically calculated 
IV-characteristic for forward bias. 
The low voltage part of all the IV-characteristics computed in Figs.(\ref{finivcamgdmult}) 
and (\ref{finivcamgapprox}) for different values of the injection barrier can be understood in the following way. For voltages 
$V<kT/e\ll-V_{BI}$ the expression for the IV-characteristic, Eq.(\ref{ivapprox}), can be approximated by
\begin{equation}
j=e \mu_s \frac{e V_{BI}}{kT}\frac{V}{L_{eff}} {\cal N} \exp \left ( -\frac{\Delta_{eff}^+}{kT} \right ),
\label{jappr1}
\end{equation}
and shows a linear dependence of the current on the voltage. 
At zero voltage and current, electrons drift from the magnesium
to the calcium electrode through the organic layer as a 
consequence of the difference in the work functions of the two electrodes. 
The blocking character of the right electrode restricts the effective electron density in the device as is seen from the exponential 
factor in Eq.(\ref{jappr1}).

For voltages from the wide range $kT/e\ll V <-V_{BI}$ but not very close to $-V_{BI}$, so that $-e(V+V_{BI})\gg kT$, the 
IV-characteristic reveals a sharp exponential increase. Such a behavior can also be extracted by an approximation of equation 
(\ref{ivapprox}) for voltages from this range (see Fig.(\ref{finivcamgapprox})),
\begin{equation}
j=e \mu_s \frac{V+V_{BI}}{L_{eff}} {\cal N} \exp\left ( -\frac{\Delta_{eff}^+}{kT} +\frac{eV}{kT} \right )   .
\label{jappr2}
\end{equation}
 The exponential increase stops when the voltage $V$ approaches the value of the built-in voltage, $-V_{BI}$.  At this point,  
the mean electric field in the device changes sign.

Nevertheless, well above the built-in voltage, so that  $V+V_{BI}>>kT/e$, both the computed characteristic and Eq.(\ref{ivapprox}) show another exponential dependence of the 
current on the voltage, due to the barrier lowering (see Fig.(\ref{finivcamgapprox})):
\begin{equation}
j=-e\mu_s \frac{V+V_{BI}}{L_{eff}} {\cal N} \exp \left( -\frac{\Delta^-}{kT} + \epsilon  \frac{l_{TF}^-}{L_{eff}}
\frac{e(V+V_{BI})}{kT} \right ).
\label{jappr3}
\end{equation}
In both equations, (\ref{jappr2}) and (\ref{jappr3}), an exponential dependence on the voltage is observed. In the second case, 
however, this dependence is much weaker due to the factor $l^-_{TF}/L_{eff}$ in the exponent. The latter exponential increase
results from the nonvanishing width of the space charge zone in the injecting electrode. The appearance of this zone is a 
consequence of our self-consistent description of the device as a whole.

After all, when the applied voltage overcomes the barrier at the injecting electrode, $\Delta^-$, the calcium contact 
can supply more charge carriers than the bulk of the organic semiconductor can transport. That is why all the current-voltage 
characteristics end up in the space charge limited regime with $j\sim V^2$ as is seen in Figs.(\ref{finivcamgdmult}) 
and (\ref{finivcamgapprox}). Particularly, when the calcium injection barrier is below the critical value of 
$\Delta_{crit} \sim 0.27\mbox{ eV}$, the system shows a  space charge limited current $ j\sim V^2$ immediately after the built-in voltage 
is crossed, which can also be seen in Fig.\ref{finivcamgdmult}.

\section{Conclusions}

In this work, we have proposed a device model capable of describing a metal/insulator/metal device under injection limited as well 
as space charge limited conditions. The problem of defining the boundary conditions at the metal/insulator interfaces was solved by 
a self consistent treatment which fully includes the metal electrodes in a consistent one dimensional description of the device. In 
this treatment boundary conditions are defined far into the bulk of the metal electrodes where the respective media is well defined. 
The values for the electric field and the charge carrier densities at the interface can be calculated and depend on the condition 
of the considered system.

We applied our model to organic semiconductors sandwiched between metal contacts, being interesting for optoelectronic applications. 
Though existing models include specific features of organic semiconductors, like energetic disorder or hopping transport, they 
describe charge-carrier injection mostly in the two limiting cases of high or low injection barriers. In the former case charge-carrier
injection models are based on the one-electron-picture within the framework of the Fowler-Nordheim tunneling model or the 
Richardson-Schottky model of thermionic injection 
\cite{Scott1999,Malliaras1999,Arkhipov1998,Arkhipov1999,Arkhipov2003,Scott2003,Preezant2003,Shen2004,Woudenbergh2005,Novikov2006}, 
where space charge effects are ignored completely. In the latter case the metal/insulator contact is assumed to be ohmic. 
Here, the contact can supply much more charge-carriers than can be transported through the bulk of the insulator and hence bulk 
properties alone define the characteristics of the system with no influence of the injection barrier 
\cite{Blom1996APL,Blom1996PRB,Neumann04}. As far as we know, the only model accounting consistently for both injecting 
barriers and space charge effects is that of Tuti\v{s} et al. \cite{Tutis2001}. The latter numerical treatment, however, does not 
allow analytical fitting of the current-voltage characteristics, making the physical reasons for the cross-over from barrier to
space-charge dominated type of device behavior inaccessible.

By contrast, the approach described here, offers the possibility to calculate IV-characteristics of metal/insulator/metal systems 
using experimentally accessible input parameters. Our model is capable to predict IV-characteristics in forward and 
reverse bias and for all values of the injection barriers at each interface. The description includes the built-in voltage and field 
dependent effective injection barriers, being direct consequences of the self consistent approach. Calculated IV-characteristics can 
be divided in different regimes for 
which approximate solutions were derived. For low voltages a linear dependence of the current on the voltage is observed, followed by 
an exponential increase. This behavior is not due to intrinsic charge carriers, but due to diffusive charge carrier transport. Once 
the voltage exceeds the built-in voltage a weak exponential current increase is observed as a consequence of injection barrier 
lowering. Here, charge carrier injection dominates the IV-characteristic. 
At very high voltages, barrier lowering becomes so strong that more charge carriers 
are injected than can be transported through the bulk of the semiconductor. This leads to the SCL dependence, 
$j \propto V^2$, for all calculated IV-curves. The relative occurrence of these different regimes thereby depends strongly on the 
injection barriers at the contact between the insulator and the anode and the cathode metals, respectively. As a consequence,  
rectifying behavior is observed for strongly disparate contacts.\\
A more realistic model incorporating charge-carrier trap states and Gaussian DOS distributions of the organic semiconductor 
is currently developed.

\section{Acknowledgements}
The authors acknowledge the Deutsche Forschungsgemeinschaft for financial support of the Sonderforschungsbereich 595.

\newpage

\bibliographystyle{plain}
\bibliography{apssamp}

\end{document}